# Measurement by FIB on the ISS: two emissions of Solar Neutrons detected?

**Y. Muraki[2], K. Koga[1], T. Goka[1]\*, H. Matsumoto[1], T. Obara[1+], O. Okudaira[1], S. Shibata[3], and T. Yamamoto[4]**

[1] Tsukuba Space Center, JAXA, Tsukuba, 305-8505 Japan

[2] Solar-Terrestrial Environment Laboratory, Nagoya University, Nagoya, 464-8601 Japan

[3] Department of Information Science, Chubu University, Kasugai, 487-8501 Japan

[4] Department of Physics, Konan University, Kobe, 658-8501 Japan

\* present address: Faculty of Engineering, Tokyo City University, Tokyo, 158-8557 Japan

+ present address: Planetary Plasma and Atmospheric Research Center, Tohoku University

Correspondence to: Y. Muraki (muraki@stelab.nagoya-u.ac.jp)

## Abstract

A new type of solar neutron detector (FIB) was launched on board the Space Shuttle Endeavour on July 16, 2009, and began collecting data at the International Space Station (ISS) on August 25, 2009. This paper summarizes the three years of observations obtained by the solar neutron detector FIB until the end of July 2012. The solar neutron detector FIB can determine both the energy and arrival direction of neutrons. We measured the energy spectra of background neutrons over the South Atlantic Anomaly (SAA) region and elsewhere, and found the typical trigger rates to be 20 and 0.22 counts/sec, respectively. It is possible to identify solar neutrons to within a level of 0.028 counts/sec, provided that directional information is applied. Solar neutrons were possibly observed in association with the M-class solar flares that occurred on March 7 (M3.7) and June 7 (M2.5) of 2011. This marked the first time that neutrons had been observed in M-class solar flares. A possible interpretation of the production process is provided.

# 1   Introduction - a brief history of solar neutron detection

High-energy protons coming from the Sun on February 28 and March 3, 1942 were first discovered by Forbush and published in 1946.[1] In 1951, Biermann, Haxel, and Schulter had predicted the potential discovery of solar neutrons on Earth.[2] Neutrons are produced when the accelerated ions strike the solar surface. However, solar neutrons were actually detected 29 years after this prediction. A clear signal of gamma rays and neutrons was detected in association with a large solar flare on June 21, 1980, with an X-ray intensity of X2.5, by the Gamma Ray Spectrometer composed of the NaI and CsI detectors on board the Solar Maximum Mission (SMM) satellite.[3] Figure 1 shows the results. The first peak corresponds to the gamma-ray signal, while the second was induced by the neutron signal. Because neutrons cannot travel from the Sun to Earth at the speed of light, their arrival time distribution is associated with a time delay from the speed of light, even when simultaneously released from the Sun. For the time distribution presented in Fig. 1, if the same departure time for neutrons is set, a neutron energy spectrum is obtained, which assumed an impulsive production of neutrons on the Sun. The spectrum can be expressed by a power law: $En^{-\gamma}dEn$ with $\gamma = 3.5 \pm 0.1$.[3]

Two years later, on June 3, 1982, the SMM satellite again detected a neutron signal.[4] However, neutron monitors located on the ground have successively detected neutron signals in association with a large X8.2 solar flare[5], which shed new light on the production time of neutrons in the solar atmosphere. One component involved in the data cannot be explained by an impulsive production mechanism alone. Only two solar neutron events had been accumulated until solar cycle 21, and it was too early to judge the production time of neutrons in the solar atmosphere, namely, whether high-energy neutrons are produced impulsively or gradually. Both scenarios were possible for the same event and this would be a great challenge for solar physicists.

To identify the production time of neutrons at the solar surface in the solar cycle 22, new detectors capable of measuring the energy of neutrons were expected. Therefore a new type of solar neutron detector - the solar neutron telescope (SONTEL) - was designed, based on the plastic scintillator. SONTEL can measure the energy and direction of neutrons using the charge exchange process of neutrons into protons[6]. Therefore SONTEL can determine the flight time from the Sun to Earth. Of course, conventional Simpson-type neutron monitors were also operated[7-8]. At the same time, the possibility of launching a new type of solar neutron detector into space was considered[9]. To resolve the mystery regarding the production time of neutrons, it is inevitable to have a new type of detector capable of measuring the energy of neutrons in space. We could then identify *when* those neutrons left the Sun.

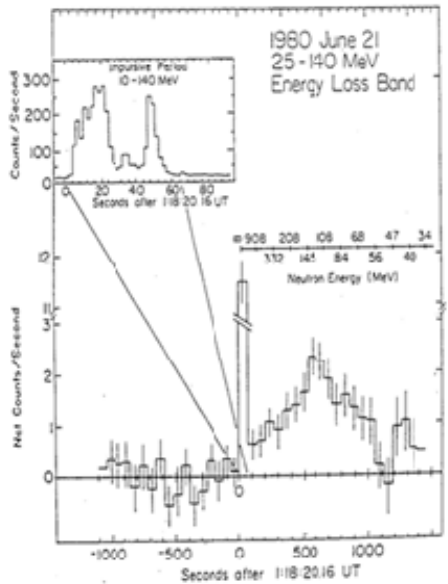

Figure 1. Solar neutrons detected on June 21, 1980. Gamma-rays accounted for the first peak; neutrons accounted for the second. The original picture was prepared by the authors of the ref [3].

An attempt to measure the energy of neutrons was found in a paper dated around 1985[10]. Scintillator bars composed of two layers were equipped (in X and Y directions) and the device was circulated over the Southern hemisphere by a Racoon balloon flight from Alice Springs, Australia to detect solar neutrons within the energy range of 20 to 150 MeV, but detected no signals. Almost the same year 1985, another new instrument was proposed, capable of unambiguously determining the energy and direction of incident neutrons using a technique, the double Compton scattering method[11]. The detector was named SONTRAC. Independent from these activities, most of which were developed in the USA, in Japan, a detector comprising a mass of the scintillation fiber is proposed to detect anti-deuterium in space and neutron and anti-neutron oscillation in the space between the Sun and Earth. In 1991, we proposed a new type of detector for the Japanese Experimental Module (JEM) of the International Space Station (ISS)[12-13].

In April 1991 and August 1991, a large gamma-ray satellite CGRO and a solar satellite Yohkoh were launched.   An image of the Sun using neutron signals was successfully drawn, using the Compton scattering function of the COMPTEL detector [14] and beautiful photographs of solar flares were taken by using the soft and hard X-ray telescope of Yohkoh satellite.   They have left very important archives on the solar activities [15].

During the solar cycle 22, several new discoveries involving solar neutrons were made; based not only on many ground level detectors but also a few that were space-borne. Consequently, more solar neutron events were accumulated, including one highlight, the discovery of an extremely strong signal of neutrons in association with the large X9.3 solar flare on May 24, 1990[16].  The signal was the strongest ever observed by the neutron monitor.   In association with this flare, two Soviet satellites, GRANAT/PHEBUS [17] and GAMMA-1[18-19] successfully captured very impulsive high-energy gamma rays starting at 20:48 UT.   One minute later (20:49UT), strong neutron signals were detected by many neutron monitors located throughout the North American continent [20].   Subsequently, from around 21:00UT, Ground Level Enhancement (GLE) was observed, induced by high-energy protons.   The key knowledge obtained by the event on May 24, 1990, reviewing some 20 years after the discovery, may be the sudden increase in the ratio between 70-95MeV gamma rays and 4-7 MeV nuclear gamma rays 3 minutes later. Chupp and Ryan summarized that the change in ratio may have been induced by accelerated protons to several hundred MeV [21].   It is worth noting that to detect high-energy neutrons at ground level, conventional neutron monitors were not only used to detect solar neutrons in the large solar flare on May 24, 1990, but also in the X9.4 flare on March 22, 1991[22].

The subsequent scope of remarkable events from solar cycle 22 may also include the detections of high-energy gamma rays and neutrons in association with the six extremely powerful solar flares with X12 observed during June 1 and 15, 1991.   Solar neutrons were detected in association with two large solar flares on June 4 and 6, 1991, respectively, using two kinds of solar neutron detectors located on Mt. Norikura: the solar neutron telescope [23] and the neutron monitor [24].  Via simultaneous observations with the neutron monitor and the neutron telescope, the capability of the new solar neutron telescope was demonstrated.

It should be mentioned that in the solar flare on June 4, 1991, the BATSE[25] and OSSE[26] detectors on board the CGRO satellite observed the long-standing emission of gamma rays with a decay time of 330 seconds after a sharp impulse signal. OSSE observed a neutron capture line (2.223 MeV) and a carbon de-excitation line (4.44 MeV) that continued for three hours. High-energy gamma rays were detected by the EGRET detector with energies of 50 to 100 MeV and >150 MeV for the flare events on June 4, 6, 9, and 11. Moreover, in the flare event on June 11, a particularly long-lasting emission of high-energy gamma rays was recorded, lasting 10 hours [27].

Many arguments concerning the long-lasting gamma rays emerged at the time, namely whether they were induced by the continuous acceleration process of protons (such as in the shock acceleration model[28]) or by protons trapped in the magnetic loop and precipitating on the solar surface.[29-30] The impulsive production mechanism of neutrons on the solar surface was attributable to the reconnection process of magnetic loops[31-33] or the DC acceleration mechanism[34], while long-lasting emissions of gamma rays may be closely related to the shock acceleration process. The question of whether the long-lasting high-energy gamma-ray emission is attributable to the continuous acceleration of the protons above 300 MeV [35], or the injection of flare accelerated particles into a large coronal loop with release at the mirror points of the loop where the gamma rays are produced, is very interesting [29-30], the final answer to which will hopefully be obtained in solar cycle 24.

During the solar flare event on September 7, 2005, solar neutron telescopes located on Mt. Sierra Negra in Mexico (at 4780 m) and Mt. Chacaltaya in Bolivia (at 5,250 m) both observed a clear solar neutron signal[36], which was also recorded by three different counters located in the Northern and Southern Hemispheres. This made it possible to compare the detection efficiency of a solar neutron telescope with that of a conventional neutron monitor. The detection efficiency ratios were found to be 1 and 0.7, for the neutron monitor and neutron telescope respectively, pertaining to the same area of both detectors. Since the solar neutron telescope cuts low-energy neutrons of less than 30 or 40 MeV, its detection efficiency is also lower than that of the neutron monitor [37]. The neutron monitor is highly sensitive to neutrons with energy exceeding about 10 MeV [8]. It is worth noting here that the data suggests the involvement of neutrons produced by both the impulsive and gradual phases [36].

The FERMI-LAT satellite also recently observed two gamma-ray events in association with M-class solar flares on March 7 and June 7, 2011[38]. Again a long duration component lasting more than 14 hours was observed and the continuous emission of GeV gamma rays from the Sun was detected. This mechanism may indicate a different mechanism in the gamma-ray production process in addition to that responsible for the impulsive production of gamma rays, which is discussed in the final part of this paper. An effective summary on solar neutron research has been recently published in a book, which also contains more detailed bibliography [39].

The aim of this paper is to present new results using the FIB detector on board the ISS. Actually Chapter 2 introduces details of the new solar neutron telescope FIB detector, followed by the neutron observation results on the ISS in Chapter 3. Chapter 4 covers the solar neutron events observed using the new detector in association with the M-class solar flares on March 7, 2011 (M3.7), and on June 7, 2011 (M2.5). Chapter 5 discusses our results compared to other observations, and Chapter 6 summarizes the results.

## 2 New Solar Neutron Detector FIB on the ISS

### 2-1 SEDA-AP-FIB detector

The new solar neutron telescope has been designed as a component of SEDA-AP. A detector for Space Environment Data Acquisition equipment - Attached Payload (SEDA-AP) was originally proposed to measure radiation levels at the International Space Station (ISS) in 1991[12-13]. In 2001, an actual Flight Module (FM) was ready to be deployed, but an accident involving the Space Shuttle resulted in the FM being stored in a special clean room for eight years until it could finally be launched.

SEDA-AP was designed as one of the detectors on board the Japan Exposure Module (JEM). This equipment not only comprises a neutron detector but also various other detectors, such as charged particle detectors, a plasma detector, an atomic oxygen monitor, and electronic device evaluation equipment. The system even includes a micro-particle capture detector.

The neutron detector consists of two parts: a conventional Bonner Ball Detector (BBD) and a FIBer detector (FIB). The name FIB is so-called because the main part of the sensor is comprised by the mass of the scintillation fiber. The BBD measures low-energy neutrons; the FIB measures high-energy neutrons. Technical details can be found on the JAXA website (http://kibo.jaxa.jp/en/experiment/ef/seda_ap).

The neutron detector can be extended 1 m from the main frame via a mast to reduce the background neutrons coming from the vessel of the SEDA-AP. The system has a 220-watt power supply and a total weight of 450 kg. The FIB was launched by the Space Shuttle Endeavour on July 16, 2009, and began taking measurements at the ISS on August 25, 2009. Since then, the detectors have been working problem-free. Although the official mission lifetime was estimated as three years, given the importance of the measurements, it would be highly desirable to extend this period to cover at least one solar cycle of 11 years, provided that the system continues to operate.

### 2-2　The experimental purposes

This experiment has three main scientific goals as follows:

(1) Accurate measurements of radiation levels in the ISS environment[40-41]

(2) Rapid prediction of the imminent arrival of numerous charged particles from the Sun by monitoring GeV GLE particles for the flares of the western part of the solar surface (space weather forecast). However for the flares of the eastern part of the solar surface, the amount of emitted high energy particles may be estimated by observing neutrons.

(3) Identification of the production time of neutrons induced by the accelerated protons above the solar surface. We wish to know *when* an*d how* high-energy particles are produced over the solar surface. When high-energy charged particles arrive at Earth and are detected, important information may be lost concerning the production time at the Sun. To understand the acceleration mechanism of charged particles at the Sun, it is necessary to compare the data of neutrons and gamma rays with images taken by a soft X-ray telescope[42], RHESSI and/or the UV telescope launched on the Solar Dynamical Observatory.[43]

To determine the neutron production time at the Sun, it is necessary to employ a neutron detector capable of measuring the energy of neutrons. Currently, no such detector has been used in space other than an FIB detector, although the ground-based Solar Neutron Telescopes (SONTEL) have been operating for a number of years.[44-46] Accordingly, the FIB detector installed in SEDA-AP may provide a crucial data measuring neutron energy in space in the solar cycle 24.

### 2-3  Sensor design, detection efficiency, and trigger

To achieve the scientific goals listed earlier, a fine-grated neutron detector FIB has been designed, consisting of a plastic scintillator with 32 layers (sheets) and dimensions of 3 mm (height) × 96 mm (width) × 96 mm (length). Sixteen stacks of scintillation bars are used per layer, with each bar having dimensions of 3 mm (height) × 6 mm (width) × 96 mm (length). Each layer is located along the x-axis and y-axis alternatively, forming a stratified block and an optical fiber is coupled to the end of each bar to collect photons produced in the scintillator. These photons are then sent to a 256-channel multi-anode-photomultiplier (Hamamatsu H4140-20). Figure 2 shows a schematic image of the FIB detector, which measures the tracks of recoil protons produced by incident neutrons and determines neutron energy using *the range method*.

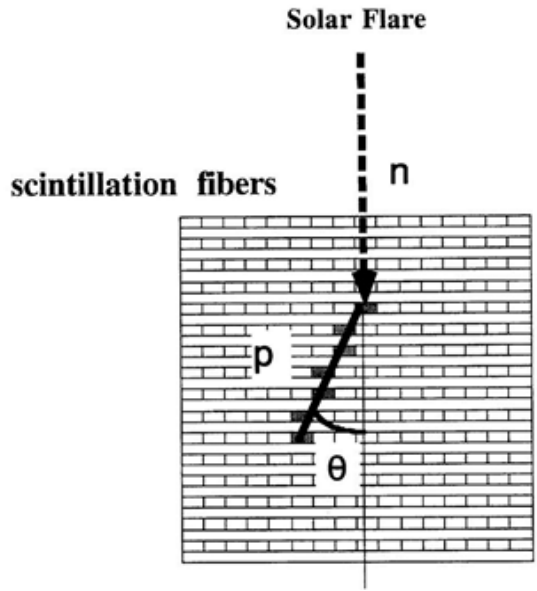

Figure 2. Schematic view of the FIB detector's sensor. One layer consists of 16 plastic scintillator bars with dimensions of 3 mm (height) × 6 mm (width) × 96 mm (length). The direction of the Sun is identified by tracking two layers of the scintillator in both X and Y directions; the energy of neutrons is measured by the range of protons.

It can also identify the direction of neutron incidence. Neutrons and protons are discriminated by an anti-coincidence system consisting of six scintillator plates surrounding the FIB sensor in a cubic arrangement. To measure the total radiation dose at the ISS, we actually collect neutron data obtained over the South Atlantic Anomaly (SAA) region. The maximum count rate of the anti-coincidence system for the SAA region is 60,000 counts per second, and it works.

The cubic-shaped sensor used for neutron detection has sides measuring 10 cm and maximum kinetic energy of about 120 MeV. As shown in Figure 3, the sensor is monitored from two directions by two multi-anode photomultipliers (PMT1 and PMT2), meaning the arrival direction of the tracks can be identified. To determine the arrival direction of neutrons, protons must penetrate at least four sensor layers, each of which consists of plastic bars 3-mm thick. Consequently, the lowest neutron energy that can be measured is 35 MeV.

A trigger signal is produced by dynode signals from the PMTs (it is set at ≥ 30MeV proton equivalent). When the dynode signals from both PMTs exceed a certain threshold, a trigger signal is produced. When the trigger rate is less than 2 counts/sec, all ADC values for each channel are recorded in memory. The analog memory can handle all 512 channels of both PMTs. When the trigger rate exceeds 2 counts/sec, only the on-off signal (1 or 0) of each channel is recorded. When it exceeds 15 counts/sec, only the total output signal of the dynode is recorded. The technical details can be found elsewhere.[48-50]

Figures 3. (a) The photograph of Japan Exposure module onboard the International Space Station. The coordinate of the FIB sensor is drawn on by the blue arrows (X coordinate blue, Y coordinate in red and Z coordinate by green), (b) An FIB sensor was mounted on SEDA-AP together with BBD. Proton tracks are measured by a photomultiplier (PMT2) that looks the scintillation fibers from the bottom side (Z-Y plane), while for the X-Z plane the other photomultiplier (PMT1) is used looking from the proceeding section of the ISS.

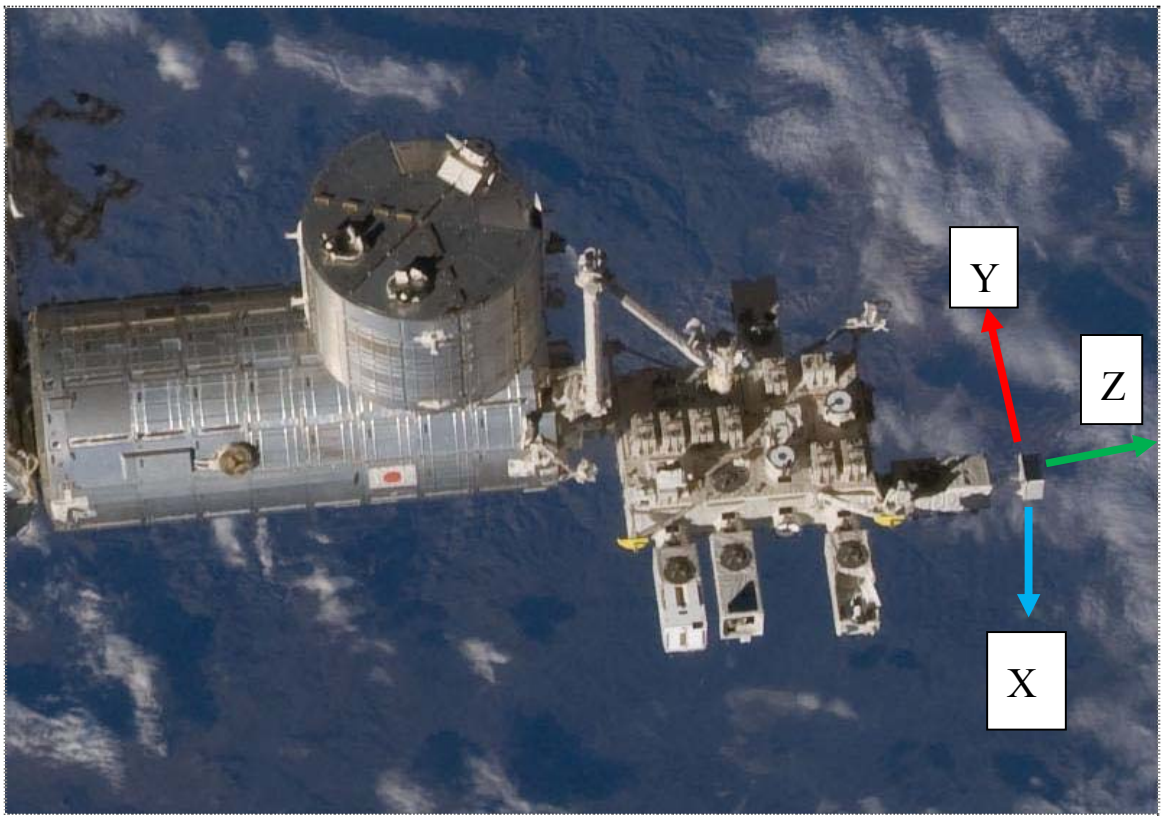

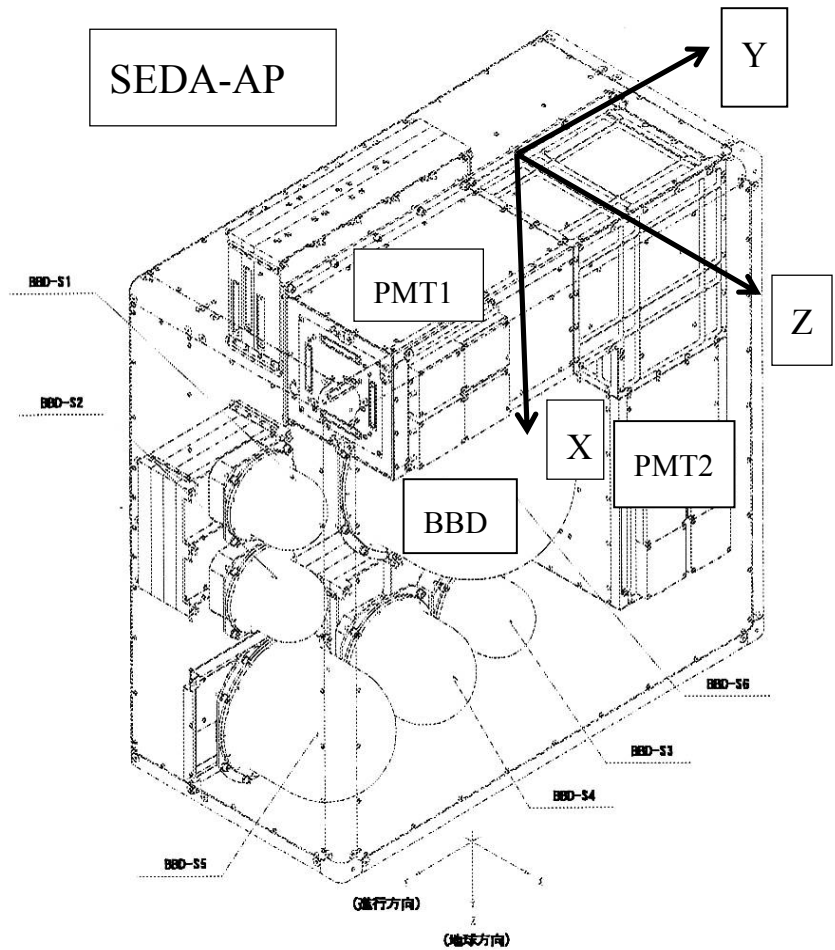

## 2-4 Detection efficiency

We now discuss the neutron detection efficiency of the sensor. Because of the cubic-shaped detector has sides measuring 10 cm, if neutrons with kinetic energy not exceeding 120 MeV interact at the top of the detector, the track of protons will be fully contained in the apparatus.　However, if n-p scattering occurs in the lower part of the detector, the recoil protons will escape by crossing one anti-counter plane. This means the anti-counter is triggered and the neutron event will not be recorded, resulting in a geometrical factor dependent on energy.　Furthermore, the nuclear interaction cross-section also depends on the energy of neutrons, which imposes a greater energy dependence on detection efficiency.

We actually obtained detection efficiency (ε) using the Monte Carlo method and the Geant4 program, during which the collisions between neutrons and the carbon target were also taken into account. The detection efficiency of neutrons for vertical incidence without using any anti-coincidence panel can be approximated by ε = 3.45 × E[MeV]$^{-0.7118}$. For example, neutrons with incident kinetic energy of 100 MeV is expected to be detected by an efficiency of 0.13 or 13%. For the incidence with θ = 10, 20, and 30º, the coefficient 3.45 is replaced by 3.1, 2.9, and 2.6, respectively. However, another condition has been applied to data analysis, whereby the minimum energy deposited in the sensor exceeds 35 MeV. According to the Monte Carlo calculation, the detection efficiency (ε) can be expressed as ε = 1.15 × (E-25[MeV]) × E[MeV]$^{-1.8}$. The results are given in Figure 4. In fact, the detection efficiency (ε) could be approximately expressed by a constant value of 0.021 (almost 2%) over a wide energy range of incident neutrons where En = 50-120 MeV. We took account of these efficiencies when obtaining the energy spectrum of neutrons.

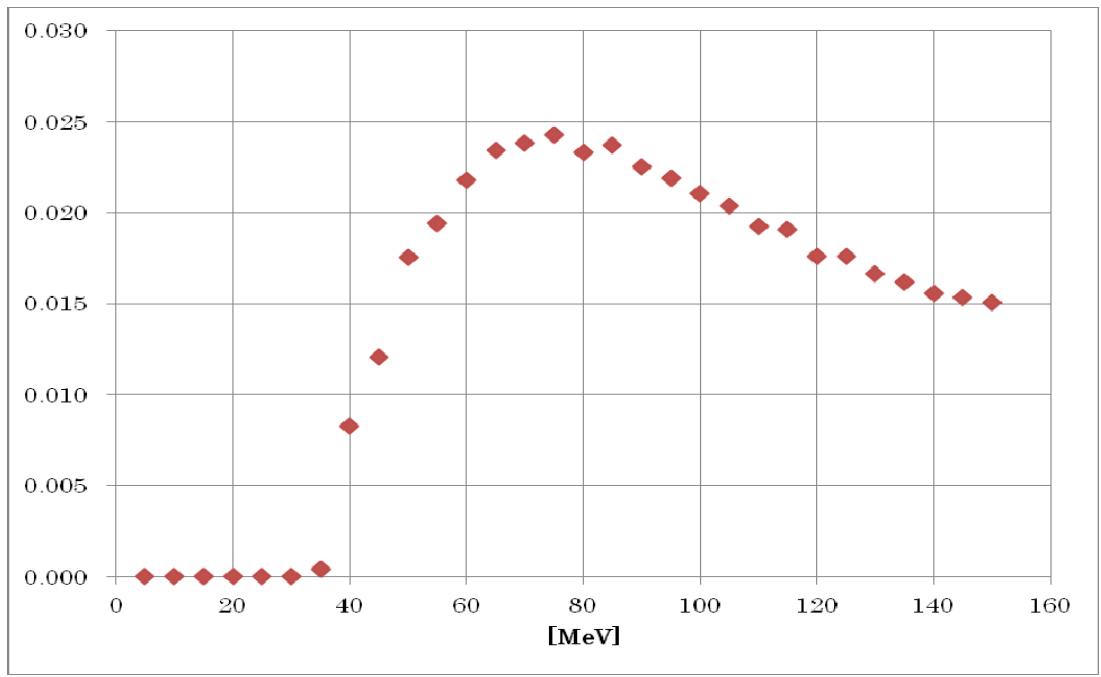

Figure 4. The detection efficiency of the FIB sensor for neutrons as a function of incident energy. The vertical value of 0.02 corresponds to the detection efficiency of 2%. The curve was obtained by the Geant 4 program, taking account of the collision processes of neutrons with the proton or carbon target inside the scintillation fibers. The anti-counter trigger condition was also taken into consideration.

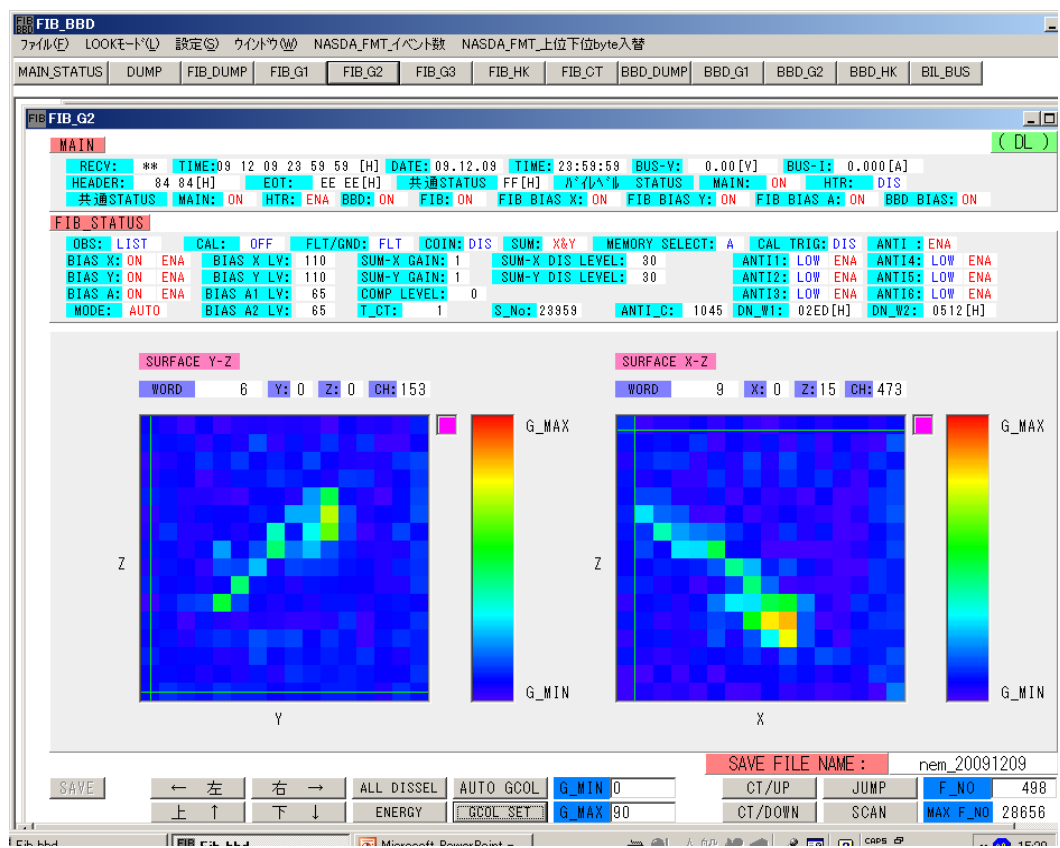

Figure 5. A typical background neutron event. The neutrons enter from the preceding section of the ISS (small to large in the horizontal scale Y of the left side picture) and from the top to the bottom (small to large of the right side picture X in horizontal axis). The incident neutrons are converted into protons inside the scintillator. The energy deposit in each bar is color-coded. At the termination point of the track, the Bragg peak can be recognized. The G_min and G_max correspond to the ADC channel 0 and 90. The linear ADC has a range of 256 channels and the dip and peak correspond to 11 (dark blue) and 22 (light blue) respectively induced by the minimum ionizing particles. Orange spot corresponds to about channel 70.

## 3 Measurement of Neutrons on the ISS

This chapter presents the actual neutron measurement results obtained on the ISS. Figure 5 shows a typical event detected by the NEM sensor. The image on the right was taken by the PMT located in the proceeding section of the ISS (X-Z plane), while that on the left is a photo taken from underneath the sensor (Z-Y plane) (i.e. looking up from Earth, See Figure 3b). The z-y sensor points upward toward Earth and the z-direction points towards the opposite side of the pressurized ISS module. The color represents the amount of energy deposited in each scintillating bar of dimensions of $6 \times 3 \times 96$ mm. Figure 6 presents the counting rate of the FIB on March 7, 2011. Each panel of Figure 6 shows from the top to the bottom, the position of ISS, the strength of the magnetic field, the counting rate per minute and the integral counting rate respectively. The satellite observed the solar flare after 20:02UT.

Figure 6. From the top to the bottom, each figure corresponds to (1) the location of the satellite; whether it was over the day side of the Earth or the night side, (2) the strength of the magnetic field measured by SEDA-AP, (3) the differential event rate per minute, and (4) the integral counting rate with Universal Time. The SUN or ECL of the top panel represents whether the satellite was on the day side (SUN) or the night side (ECL) of the Earth respectively. The satellite observed the Sun after 20:02UT. Every 45 minute, the satellite passes over the Equator and approaches both Polar Regions. The FIB is an omnidirectional sensor and detects neutrons from all directions. The peak during 20:06-20:07UT may be induced by solar neutrons.

# FIB Count Profile
## (2011/03/07-19:30:00 - 2011/03/07-20:20:00)

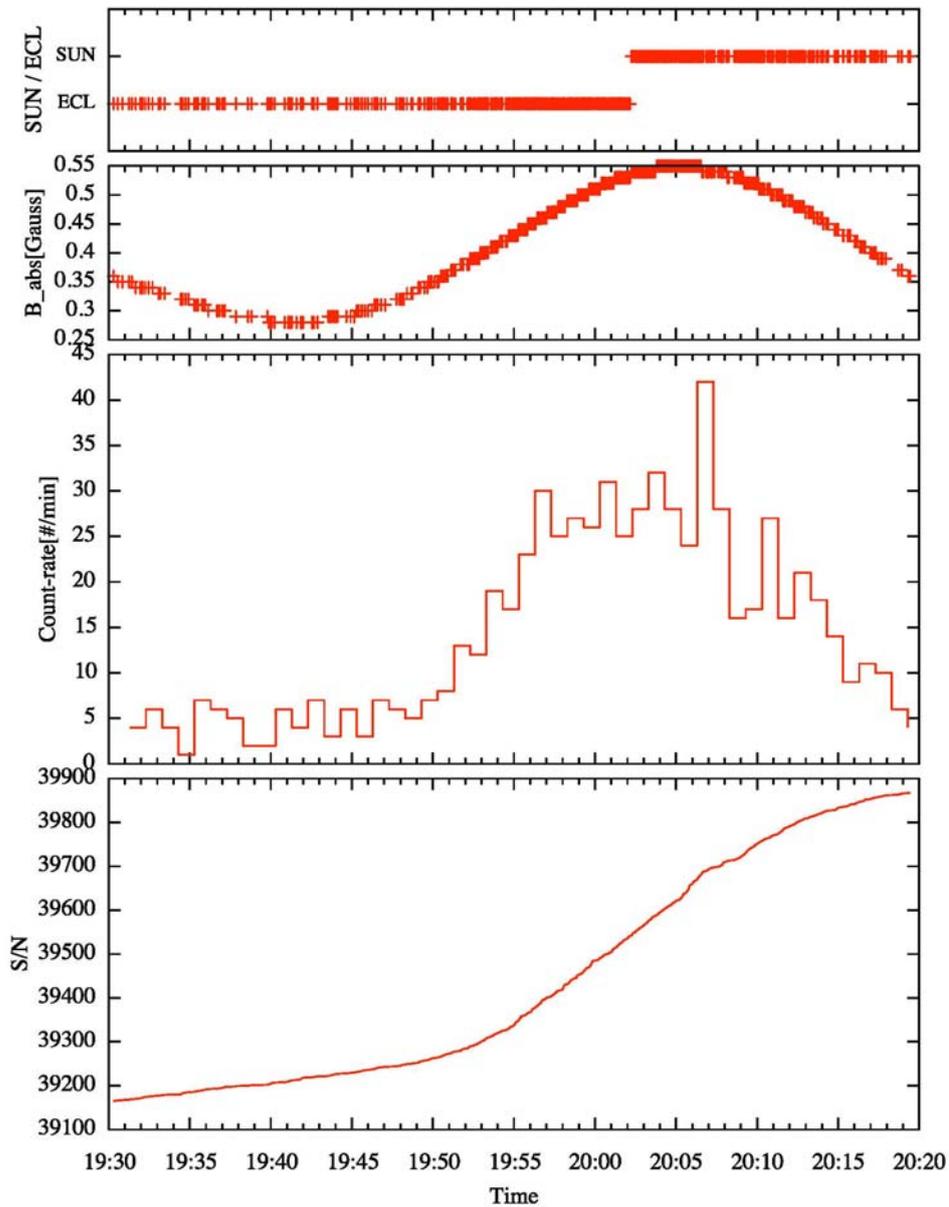

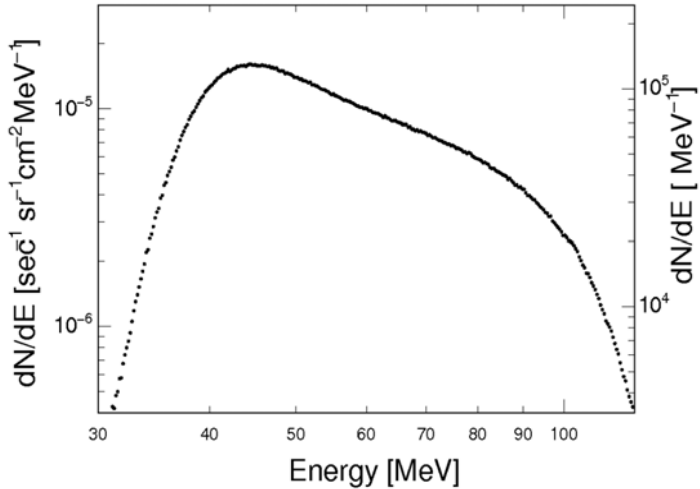

Figure 7. Energy spectrum of protons recorded in the FIB sensor during January 1 - July 31, 2010. The data obtained over the SAA region is unrelated to this data. The left side axis represents the differential flux of neutron-induced protons denoted by the unit of [/sec.sr.cm$^2$.MeV].

As evidence that the FIB detector has been working stably, the energy spectrum of neutron-converted-protons is given in Figure 7. The graph has been made by analyzing all data collected during January 1, 2010 and July 31, 2010. The observed proton spectrum can be expressed by a power law with a differential power index of -1.75 within the proton energy range where Ep = 45-85MeV. The data taken over the SAA region were excluded.

The trigger rate for neutrons was 0.22 counts/sec on average (see Figure 6 the third panel) and 20 counts/sec over the South Atlantic Anomaly (SAA) region, which is about 90 times greater than anywhere else. The orbital average value 0.22counts/sec is obtained after excluding the counting over the SAA. We have also measured neutron energy spectra using the range method. This may be the first time the energy of neutrons in space has been measured *using the range method*. As the trigger system occasionally encounters problems with memory saturation over the SAA region, it began recording only on-off data for each channel. At the end of this chapter, we mention the possibility of electron detection by this sensor. Using information on ionization loss (-dE/dX), each of the electrons is separated from the neutron-converted-protons. The internal sensor can detect electrons with energy of between 2.5 and 30 MeV, as a thin track. Figure 8 shows a candidate electron track.

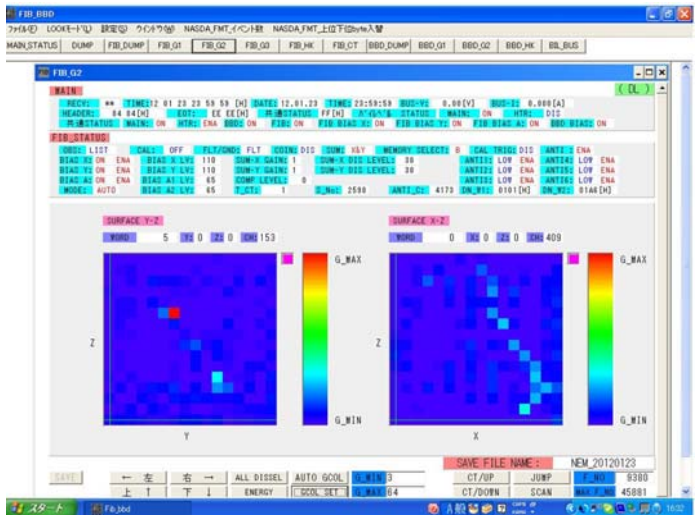

Figure 8. A candidate electron track. A photon enters from the top of the sensor (Y-Z plane). An electron apparently deposited energy inside one vertical bar (where the red spot denotes a large deposit of energy) while running vertically, thus enabling the recording of a long thin track in the X-Z plane. Since the ionization loss of electrons is smaller than that of protons, thin tracks are expected.

## 4 Solar Neutrons Associated with M-class Solar Flares

### 4.1 Search for solar neutrons by the FIB detector

According to the calculations of Imaida[47] and Watanabe[51], the typical event rate induced by solar neutrons is expected to be within the range 10 to 1,000 counts/sec in FIB. As mentioned earlier, the background rate is as low as 0.22 counts/sec, making it possible to detect all solar neutrons exceeding this level. Between September 2009 and February 15, 2011, no solar flare occurred with intensities exceeding the X-class. However, during the period February 6 to 8, 2010, four large M-class solar flares were observed, hence we analyzed the NEM data recorded at the time. Fortunately, during all four peaks of X-ray intensity, the satellite was flying over the daylight side of Earth. The peaks in X-ray intensity were observed by the GOES satellite on February 6 at 18:59 UT (M2.9), February 7 at 02:34 UT (M6.4), February 8 at 07:53 UT (M4.3), and again on February 8 at 13:47 UT (M2.0). Although we carefully searched the NEM records, there was no evidence of neutrons arriving from the Sun during any of these periods.

## 4.2 List of flares and search conditions

We searched solar neutrons for every solar flare of GOES X-ray intensity exceeding M2, the results of which are summarized in Table I.　The first to third columns correspond to the event date, peak X-ray intensity, and flare class, respectively. The fourth column indicates the ISS location, namely whether on the night (X) or day (○) side of Earth.　The fifth column indicates whether solar neutrons are involved in the data.　The ○→ X and X→○ notations in the fourth column indicate that the ISS was moving from the sunny side to the occultation side or vice versa 30 minutes from the peak flare time.　A ? mark in the fifth column indicates a possible neutron signal with statistical significance less than $3\sigma$ .

Thanks to past observations, neutrons are known to be typically produced when the X-ray intensity peaks, and are observed within 30 minutes of this time via an X-ray detector.[3, 4, 23, 24,36]　Solar neutrons with energy of 35 MeV need 23 minutes more than light to travel from the Sun to Earth.　Of course, the maximum time observed by the GOES X-ray detector does not always correspond to the neutron production time by the accelerated protons. Neutron energy of 35 MeV corresponds to the minimum energy for tracking protons in the FIB detector.　By taking these conditions into account, we set a data analysis time of 30 minutes and then searched for neutrons coming from the Sun.

## 4.3 Background

Before introducing the actual neutron events, let us briefly describe the background. As shown in the third panel of Figure 6, when the ISS approaches the northern or southern polar regions from the Equator, the neutron counting rate of the FIB detector increases from 0.04 to 0.5 counts/sec. The ISS completes an orbit around Earth every 90 minutes. Therefore, should solar neutrons arrive when the ISS passes over the Equator, high quality data are obtained. However, it also passes over either side of the polar regions during each 45-minute period.

For this period, the FIB detector demonstrates new functions to discriminate background and squeeze signals. We assume that the direction of *protons induced by solar neutrons* was observed within a cone with an opening angle of about 45 degrees relative to the direction of the Sun. By applying a simple acceptance calculation ($1/2*\pi$ steradian / $4\pi$ steradian = 1/8), the background may be reduced to 1/8. We can therefore identify neutron signals under a background level of 0.028 counts/sec (= 0.22/8). In other words, when the intensity of solar neutrons is weaker than 0.028 counts/sec, solar neutrons will become far more difficult to detect.

## 4.4 Actual event observed on March 7, 2011

On March 7, 2011, in association with the M3.7 flare, a possible signal of solar neutrons was captured by the FIB detector and more than 54 proton events were identified as coming from the direction of the Sun. In Figure 9, we present a distribution of the arrival direction of those 54 events over the background at the same time.

The statistical significance of the event was 6.8σ (based on the Li-Ma method). We regard those protons as being produced by solar neutrons inside the FIB sensor. Figure 10 shows an example of actual event involving "solar" neutrons. The direction of the Sun was to the lower-left side for pictures on the left (Y-Z coordinate) in Figs. 10, while the solar direction was to the lower-center side for pictures on the right (X-Z coordinate). The track in Fig.10 corresponds to proton energy of 44 MeV. The event was observed at 20:10:26UT and departed time from the Sun is estimated as to be 19:50:38UT.

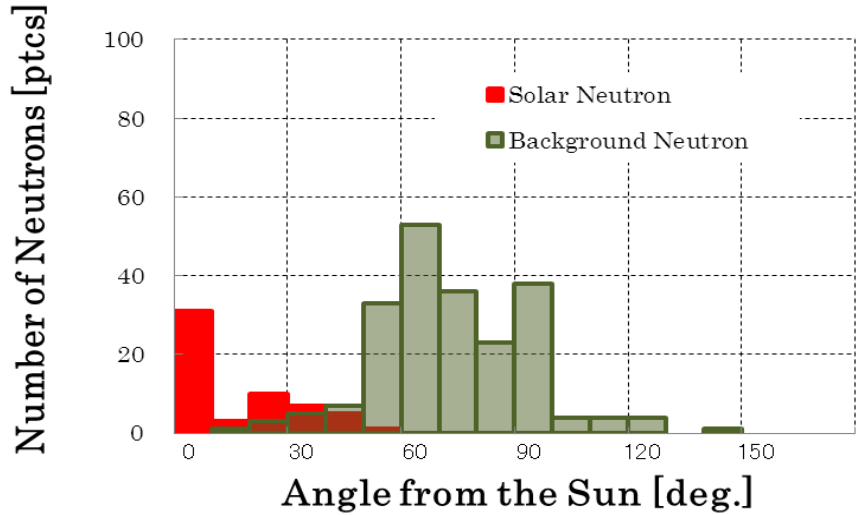

Figure 9. (a) The arrival directions of neutrons, as observed by the FIB sensor during the period 19:59:43 - 20:16:34 UT.   Neutrons coming from the Sun (red) are shown separately from the background neutrons (dark green), but those neutrons are actually observed together by the FIB sensor.   The number of events recorded during the above period was 364, and for 44, it was difficult to determine the arrival direction due to low energy. A clear peak of neutrons can be identified within the region 0 - 20 degrees from the solar direction. Meanwhile, the candidates between 20-40 degrees may be neutrons from the Sun, induced by n-C scattering and also distributed at a relatively wide angle due to the limited angular resolution of the FIB sensor for the lowest energy neutrons (about 35 MeV).   The bump of the background within the region 40-100 degrees from the solar direction corresponds to the background neutrons produced in the material of JEM by galactic cosmic rays. (b) The same data as used in the plot of (a).   However the data were normalized by the solid angle. A sharp peak can be recognized for neutrons from the solar direction (red), while the background neutrons distribute in different angles presented by the green curve.   The vertical value represents $dN/d\Omega$.   Total number of events during 20:02-20:16UT was 273 and 54 for the background and for the signal respectively.

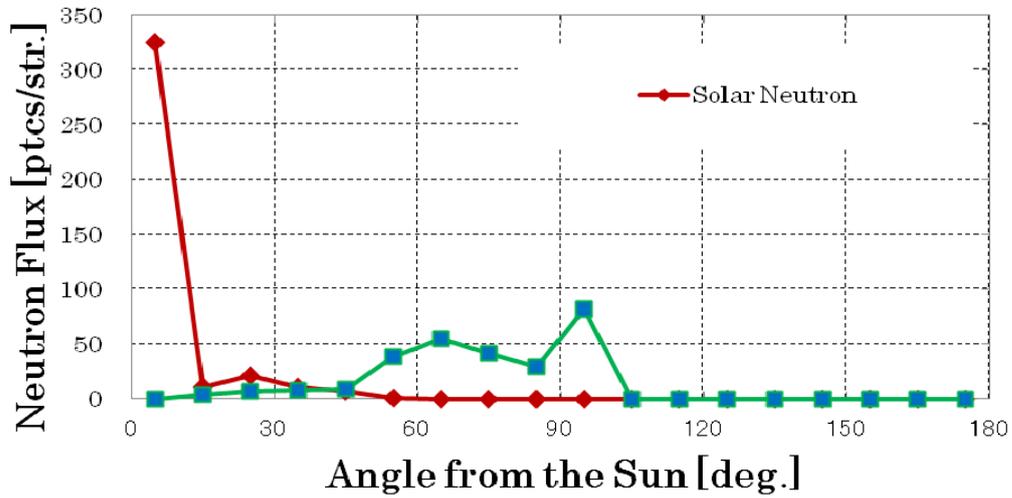

## Distribution of Incident Directions

Neutron Flux [ptcs/str.] vs. Angle from the Sun [deg.]

Solar Neutron

MAIN_STATUS | DUMP | FIB_DUMP | FIB_G1 | FIB_G2 | FIB_G3 | FIB_HK | FIB_CT | BBD_DUMP | BBD_G1 | BBD_G2 | BBD_HK | BIL_BUS

FIB_G2                                                                          ( DL )

MAIN
RECV: **    TIME: 11 03 07 23 59 59 [H]  DATE: 11.03.07  TIME: 23:59:59  BUS-V: 0.00[V]  BUS-I: 0.000[A]
HEADER:  84 94[H]    EOT: EE EE[H]    共通STATUS  FF[H]  A'(L)'S  STATUS  MAIN: ON  HTR: DIS
共通STATUS  MAIN: ON  HTR: ENA  BBD: ON  FIB: ON  FIB BIAS X: ON  FIB BIAS Y: ON  FIB BIAS A: ON  BBD BIAS: ON

FIB_STATUS
OBS: LIST    CAL: OFF    FLT/GND: FLT    COIN: DIS    SUM: X&Y    MEMORY SELECT: B    CAL TRIG: DIS  ANTI: ENA
BIAS X: ON  ENA    BIAS X LV: 110    SUM-X GAIN: 1    SUM-X DIS LEVEL: 30    ANTI1: LOW ENA  ANTI4: LOW  ENA
BIAS Y: ON  ENA    BIAS Y LV: 110    SUM-Y GAIN: 1    SUM-Y DIS LEVEL: 30    ANTI2: LOW ENA  ANTI5: LOW  ENA
BIAS A: ON  ENA    BIAS A1 LV: 65    COMP LEVEL: 0                            ANTI3: LOW ENA  ANTI6: LOW  ENA
MODE: AUTO    BIAS A2 LV: 65    T_CT: 1    S_No: 39761    ANTI_C: 2018  DN_V1: 0240[H]  DN_V2: 0146[H]

SURFACE Y-Z                                      SURFACE X-Z
WORD  11  Y: 0  Z: 0  CH: 153                     WORD  4  X: 0  Z: 0  CH: 409
                                    G_MAX                                            G_MAX

Z                                                Z

                                    G_MIN                                            G_MIN
        Y                                                X

SAVE FILE NAME:  NEM_20110307
SAVE    ← 左  右 →  ALL DISSEL  AUTO GCOL  G_MIN 3    CT/UP    JUMP    F_NO  27864
        ↑ 上  下 ↓  ENERGY     GCOL SET   G_MAX 64   CT/DOWN  SCAN    MAX F_NO 30693
        CT/UP 10   CT/DOWN 10   CT/UP 100   CT/DOWN 100

Figure 10    Solar neutron the event number 39,761 detected by the FIB sensor at 20:10:26 UT on March 11, 2011.    The solar direction was located toward the lower left side in Y-Z plane (the left side panel), while in the right side panel the solar direction was down ward almost vertically.    The energy of the track is estimated as (44±5) MeV and the neutron may be emitted around 19:50:38UT at the Sun, if the track was induced by the n-p collision. .

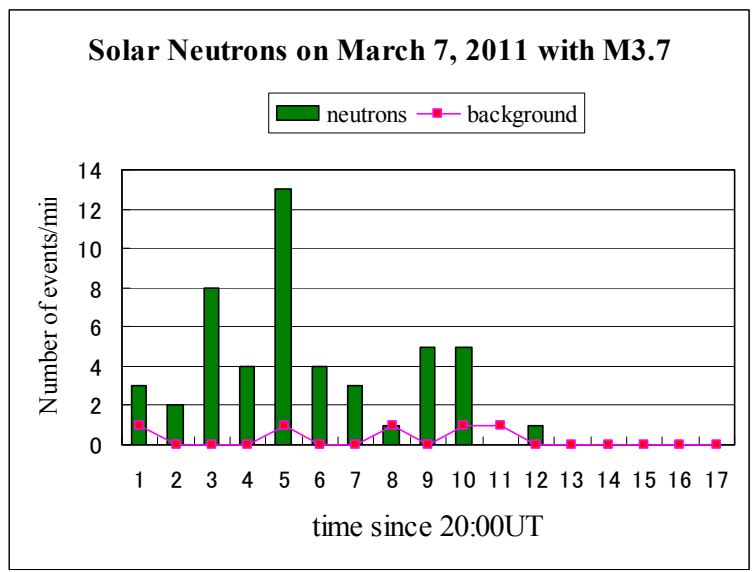

Figure 11. The arrival time distribution of neutron-induced-protons. The horizontal axis represents minutes after 20:00 UT.    The green histogram corresponds to neutrons from the solar direction.    The red curve represents the result, with equivalent analytical conditions applied to the data obtained exactly 90 minutes later.

Figure 11 presents the arrival time distribution of neutron-induced-protons. The horizontal axis represents minutes after 20:00 UT.    The red curve represents the result, with equivalent analytical conditions applied to the data obtained exactly 90 minutes later. As can be seen, contamination from the background to the possible signal involved time is relatively small.    Figure 11 also involves important information.    The final candidate of neutrons arrived near Earth 11 minutes later than the first one. The energy of neutrons was 44MeV, so we can estimate the latest departure time from the Sun to be after 19:52:36UT.

The FIB sensor has a function of measuring the energy of neutron-induced-protons. Therefore the flight time of neutrons can be estimated. The result is shown in Figure 12. In making Figure 12, the events are used that were emitted in the forward cone with an opening angle of less than 20 degrees. Figure 12 suggests that neutrons were emitted from the Sun during 19:41 and 19:54UT. It is worthwhile to note that the highest channel (7-20MeV) of the RESSHI satellite observed an enhancement during 19:42 - 20:05UT. The time of peak intensity of the hard X-rays observed by the RHESSI satellite (50-100keV) for this flare was 20:02UT.

Figure 12. The departure time distribution of solar neutrons. The histogram has been made being based on neutrons detected within a cone angle of 20 degrees from the solar direction. From the energy, the time of flight was calculated and converted the value into the departure time of neutrons from the Sun. There seem two bumps around 19:48UT and 19:52UT.

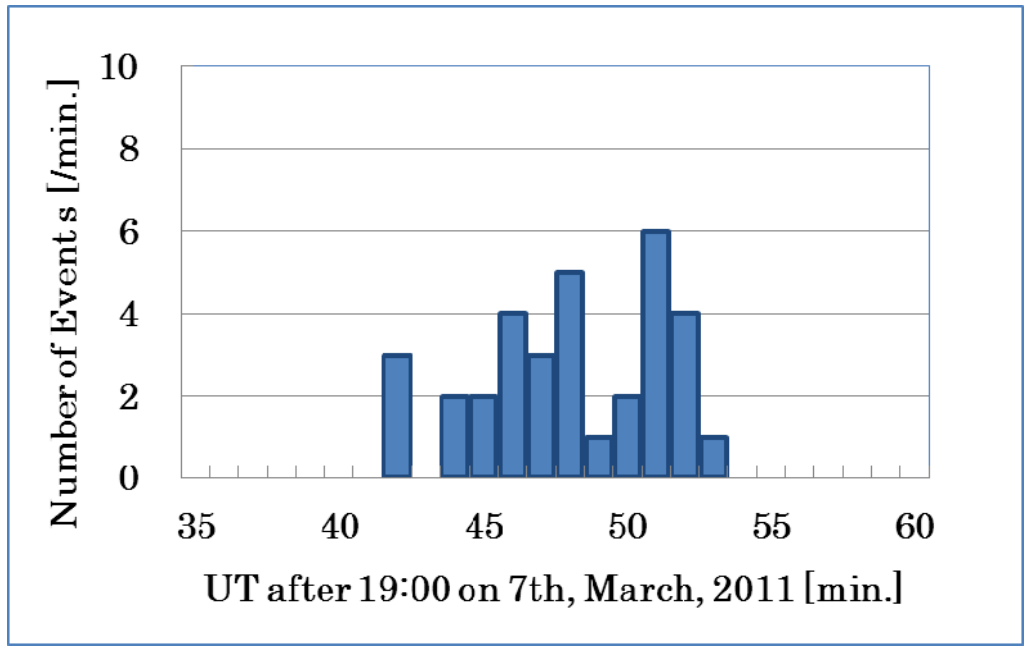

## 4.4 An interpretation on the Time of Flight distribution of neutrons

The neutron-induced-protons of Figure 12 do not necessarily reflect correct flight time of neutrons from the Sun. There is a possibility that the first bump around 19:48 UT was induced by the n-C scattering. Therefore we present in the Ep/En distribution Figure 13. To make this figure an assumption is made that those neutrons were produced at the same time on the solar surface at 19:52 UT. We could then evaluate their energy En from the flight time to the top of the atmosphere. Ep can be obtained for each neutron-induced-proton using the range method by the FIB sensor itself. Interestingly, about more than two-thirds of incoming neutron energy is converted into protons, which suggests that some of the break-up neutrons of the carbon target induced by neutrons may have escaped from the sensor to outside. This assumption has been supported by our Monte Carlo calculation based on the Geant-4 program. According to the MC calculation, the mean proton energy $<Ep>$ observed by the FIB sensor can be written by the inverse square root of the incident energy of neutrons En; $<Ep> = 0.55*En/Sqrt(En/100MeV)$. For example neutrons with incident energy of 100 MeV will be observed as protons with an average energy of 55 MeV in the FIB sensor. However we must note here since the event number of the dip between 19:49-19:50UT is not enough, so this dip has not been confirmed statistically.

Taking account of these facts; (1) the angular distribution presented in Figure 9, (2) the production time distribution of neutrons (presented in Figure 12) coincides with the duration of the flare start time (19:43UT) and the peak time (20:12UT) (measured by the GOES X-ray detector), (3) the coincidence of the flight time of the last event with an assumption that they were produced around 19:52 UT, and (4) no such enhancement was observed in the data after 90 minutes later, it would be the most natural to think this event as following: *those neutrons were produced at the Sun in association with the M3.7 solar flare.*

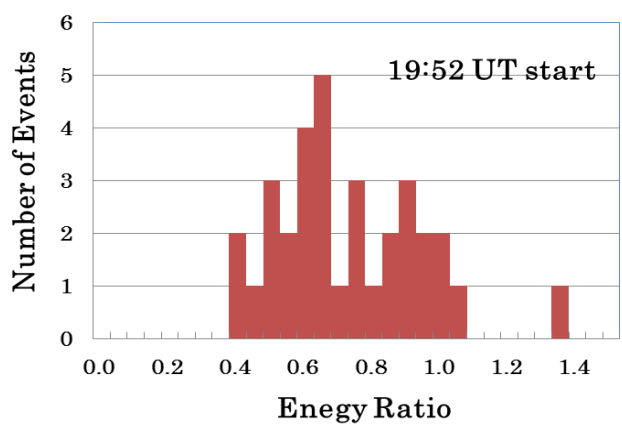

Figure 13. Energy ratio between converted protons (Ep) versus incident neutrons (E$_n$). Proton energy (Ep) was measured using the range method by the FIB sensor, while neutron energy was estimated by flight time. We assumed that the solar neutrons departed at the same time (19:52 UT) from the Sun. There is an event beyond 1.0. However taking account of the energy resolution of the sensor, it sometimes happens.

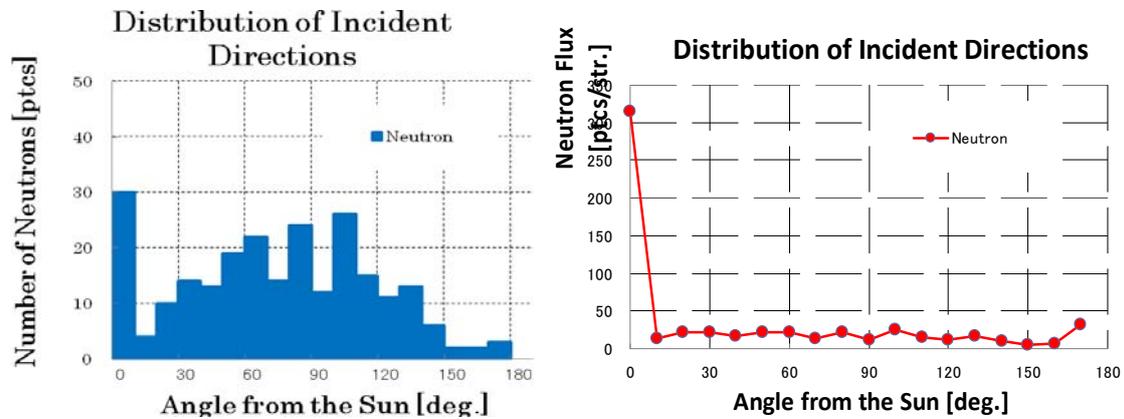

Figure 14. (a) The angular distribution of neutrons detected by the FIB sensor during 06:20UT and 06:41UT on June 7, 2011 (the left side picture). Total number of neutrons detected above duration was 240. (b) The histogram of (a) is divided by the solid angle (the right side figure). The vertcical axis represents $dN/d\Omega$. A sharp peak is recognized at the solar direction, while beyond 10 degrees, almost flat distribution is seen.

## 4.5 Other events

Three months later, the FIB detector observed another neutron event. This time the flare's intensity was only M2.5, hence far below the X-class scale. A total of 36 neutron events were identified. The statistical significance of the event was 5.8σ. The enhancement was observed during 06:21 and 06:41UT. Figure 14 shows the angular distribution of those events from the solar direction. On this event, the direction of the Sun was the opposite side of the pressurized module of ISS. So the identification from the background was easily made. The discrimination was also made by confirming the Bragg peak.

We observed three more candidate solar neutron events up to the end of July 2012: the flare events on Sep. 24, 2011 (M3.0), Nov. 30, 2011 (X1.9), and Jan. 23, 2012 (M8.7). As few as 27 events were observed on September 24, compared to the 42 events on Nov. 30, 2011, while a total of 50 neutrons were observed in association with the flare event on Jan. 23, 2012.    Those results are under preparation and will be published in elsewhere. Table II summarizes those observation results, along with other observations.

## 5   Discussions and Comparisons with Other Observations

A solar neutron event was observed on March 7, 2011, followed by another on June 7, 2011.    Both provided a new perspective regarding the production process of solar neutrons. To date, neutrons have been observed on the occasion of strong solar flares with X-ray intensity of $\geq$ X10[52].    However, the present results indicate that *even solar neutrons are produced by M-class solar flares.*    The difference in X-ray intensity between X10 and M2.5 is about 40 times greater.    The present results thus suggest that solar neutrons including M-class solar flares require a careful and complete search.

The SONG detector on board the Russian satellite CORONUS-F actually observed solar neutrons on three occasions in the solar cycle 23.    However, the solar neutrons were only detected for X-class solar flares on August 25, 2001, October 28, 2003 and November 4, 2003[53] respectively.    Moreover, neutron monitors or solar neutron telescopes observed no solar neutrons from August 2009 to the end of July 2012, which may suggest that the soft X-ray flux measured by the GOES satellite does *not necessarily* correspond to the intensity of solar neutrons from the Sun.    Although the link between this fact and the magnetic field structure[54] remains unclear, it remains a fascinating subject to be studied.

Another surprising fact is for such medium-class solar flares as those that occurred on March 7 (M3.7) and June 7, 2011 (M2.5), the LAT detector on board the Fermi satellite had observed long-lasting gamma-ray emissions.    The emission of gamma rays with energies of 100 MeV to 1 GeV continued for more than 14 hours[38].    This marked a new discovery as the EGRET detector on board the CGRO satellite observed such long-lasting gamma-ray emissions with energy exceeding 150 MeV for the flare event on June 11, 1991[35].    The RHESSI satellite observed the flare event on March 7 from its start time to the peak time, although the time profile of hard X-rays showed no unusual features and had a normal shape. The telescope of the Solar Dynamical Observatory (SDO) observed a very interesting feature; Coronal Mass Ejection (CME) started before the flare observed by the GOES/RHESSI satellites. A prominent injection of hot plasma into the base of the CME via loops was also observed by the SDO telescope. The signal of neutrons detected by the SEDA-FIB might be produced after that at the top of the inverse U-shaped loop around 19:58 UT[28], while the long-lasting high-energy gamma rays of FERMI-LAT may be produced by the precipitation of the accelerated protons on the solar surface. The activity continued for more than 10 hours[38].

For the solar flare event on June 7, 2011, hard X-ray data are available from the Fermi-GBM and RHESSI detectors. Once again, the telescope of the Solar Dynamical Observatory took a very interesting picture of this flare. The UV detector (sensitive to 171 nm) observed the very large-scale precipitation of plasma onto the solar surface[43]. In coincidence with this precipitation, protons trapped inside the upper "plasma bag" may precipitate over the solar surface. What resembles a "blow brush" over the reconnection point emits the high-energy protons confined in the plasma bag onto the solar surface, at which time the long-lasting gamma rays may be produced. However, the neutrons observed by the FIB detector may be produced when the protons accelerate, when magnetic loops are reconnected on top of the solar surface rather impulsively.

The UV telescope of SDO, however, did not observe any precipitation of plasma for the flare event on March 7, 2011, but instead a clear picture of the beginning of Coronal Mass Ejection (CME). After the *seed* of CME had been ejected, all loops under the CME started shining brightly. The protons were probably accelerated and neutrons possibly produced in one such loop via loop-to-loop interactions[55] when those protons struck the solar surface.

## 6  Summary

A new solar neutron detector was launched on July 16, 2009, on board the Space Shuttle Endeavour, and began operation at the ISS on August 25, 2009.   The sensor can determine both the energy and arrival direction of neutrons.   We measured some signals of solar neutron events in association with M-class solar flares that occurred on March 7 and June 7, 2011.   *It is interesting to note that even for M-class solar flares, we had better search solar neutrons by using both the satellite and ground level detectors.*

The precipitation of plasma bubbles, the long-lasting emission of gamma rays, and the starting of CME seeds were among the interesting features observed together with these flares.   The new data obtained certainly provides us with a new perspective on the process behind the production of gamma rays and neutrons at the Sun.

The FIB detector on board the ISS observed three further solar neutron events on Sep. 24, 2011 (M3.0), Nov. 30, 2011 (X1.9), and Jan. 23, 2012 (M8.7).   More in-depth analysis is now being conducted in connection with other data. In Table III, we compared the March 7 and June 7, 2011 events with the solar neutron event on June 21, 1980.

## Acknowledgements

The authors acknowledge the crew of the Space Shuttle Endeavour who successfully mounted the SEDA-FIB detector on the ISS Kibo exposed facility. We also extend thanks to the members of the Tsukuba operation center of Kibo for taking the SEDA-FIB data every day. The authors also acknowledge Dr. Satoshi Masuda of Solar-Terrestrial Environment laboratory of Nagoya University and Prof. Masahiro Hoshino of the University of Tokyo for valuable discussions on the solar flares of March 7 and June 7, 2011 and the acceleration mechanism of protons over the solar surface respectively.

## References

[1] S. E. Forbush, Phys. Rev. 70 (1946) 771-772. It was found that three unusual cosmic-ray intensity increases due to charged particle from the Sun. Therefore current Solar Energetic particles (**SEP**) were called as Solar Cosmic Rays (**SCR**).

[2] V. L. Biermann, O. Haxel and A. Schulter, Z. Naturforsch., 6a (1951) 47-48.

[3] E. L. Chupp, D. J. Forrest, J. M. Ryan, J. Heslin, C. Reppin, K. Pinkau, G. Kanbach, E. Riegar, and G. H. Share, ApJL, 263 (1982) L95-L98. D.J. Forrest, E.L. Chupp, J.M. Ryan, C. Reppin, E. Rieger, G. Kanbach, K. Pinkau, G. Share, and R. Kinzer, Proceedings of the 17th International Cosmic Ray Conference, 1 0, p. 5 (1981)., D.J. Forrest, E.L. Chupp, J.M. Ryan, C. Reppin, E. Rieger, G. Kanbach, K. Pinkau, G. Share, and R. Kinzer, Proceed. the 17th ICRC(Paris), 10 (1981) 5-9. The power index 3.5 is referred from the bibliography of ref. [51] by Kyoko Watanabe.

[4] E. L. Chupp, H. Debrunner, E. Flückiger, D. J. Forrest, F. Golliez, G. Kanbach, W. T. Vestrand, J. Cooper , and G. H. Share, ApJ, 318 (1987) 913-925.

[5] H. Debrunner, E. O. Flückiger, J. A. Loclwood, and D. J. Forrest, Proceed. 18th Int. Conf. on Cosmic Rays, Bangalore, India, 4 (1983) 75-78., Yu. E. Efimov, G. E. Kocharov, and K. Kudela, Proceed. 18th ICRC, 10 (1983) 276-278., Even in this flare, neutron decay electrons were observed; P. Evenson, P. Meyer, and K. R. Pyle, ApJ, 274 (1983) 875-882., D. Ruffolo, W. Dröge, and B. Klecker, AIP conference proceedings, 374 (1995) 118-123.

[6] S. Shibata, K. Murakami, Y. Muraki, M. Miyazaki, T. Takahashi, T. Sako, N. J. Martnic, and J. N. Capdeville, Proceed. 22nd ICRC (Dublin), 3 (1991) 788-791. SONTEL was named by R. Bütikofer at first. R. Bütikofer, E. O. Flückiger, L. Desorgher, M. R. Moser, Y. Muraki, Y. Matsubara, T. Sako, H. Tsuchiya and T. Sakai, Proceed. the 28th ICRC (Tsukuba), 7 (2003) 4189-4192.

[7] J. A. Simpson, Space Science Review, 93 (2000) 11-32., K. R. Pyle and J. A. Simpson, Proceed. 22nd ICRC (Dublin), 3 (2000) 53-56.

[8] M. A. Shea and D. E. Smart, Space Science Review, 93 (2000) 229-262., J. Clem and L. I. Dorman, Space Science Review, 93 (2000) 335-359., P. H. Stoker, L. I. Dorman and J. M Clem, Space Science Review, 93 (2000) 361-380., R. Pyle, Space Science Review, 93 (2000) 381-400.

[9] The idea was proposed in a symposium at Nagoya University held on April 17 and July, 1989. Also presented in the symposiums organized by JSUP of NASDA.

[10] R. Koga, G. M. Frye Jr, A. Owens, B. V. Deneby, O. Mace, and Thomas Jenkins, Proceed. 19th ICRC (La Jolla), Vol. 4 (1985) 142-145.

[11] R. S. Miller, J. R Macri, M. L. McConnell, J. M. Ryan, E. Flueckiger and L. Desorgher, Nucl. Instr. Meth. A 505 (2003) 36-40.

[12] The report to NASDA from an advisory committee of JSUP in September 1992, Junichiro Shimizu, Proceed. the 19th workshop on Space Station Utilization held on July 1-2, (1997) page 8.32-8.55.

[13] Chikaoka, R., Minutes of the 12th Space development, Committee of the Ministry of Education, Science and Culture (chaired by the Minister R. Chikaoka), Report Nos. 13-1 and 12-3, 1997. (The committee met on 9 April, 1997).

[14] M. McConnell et al., the COMPTEL collaboration, Adv. Sp. Res., 13 (1993) 245-248., M. McConnell, AIP conference Proceeding 294 (19939 21-25., J. M. Ryan and M. M. McConnell, AIP Conference Proceedings edited by R. Ramaty and N. Mandzhavidze and Xin-Min Hua, 374 (1995) 200-209., G. Rank, K. Bennett, H. Bloemen, H. Debrunner, J. Lockwood, M. McConnell, J. Ryan, V. Schonfelder and R. Suleiman, AIP conference proceeding edited by Ramaty, Manddzhavidz and Hua, 374 (1995) 219-224., J. M. Ryan, Space Science Reviews, 93 (2000) 581-610.

[15] The Yohkoh HXT/SXT flare catalogue, issued by Montana State University and the Institute od Space and Astronautical Science, edited by J. Sato, M. Sawa, K. Yoshimura, S. Masuda and T. Kosugi, published in February 2003.

[16] M. A. Shea, D. F. Smart and K. R. Pyle, Geophys. Res. Lett., 18 (1991) 1655., Proceed. 22nd ICRC (Dublin), 3 (1991), 57-60., Y. Muraki and S. Shibata, AIP Conference Proceedings, 374 (1995) 256-264.

[17] G. E. Kocharov, L. G. Kocharov, G. A. Kovaltsov, M. A. Shea, D. F. Smart, T. P. Armstrong, K. R. Pyle and E. I. Chukin, Proceed. 23$^{rd}$ ICRC (Calgary), 3 (1993) 107-110.

[18] V. V. Akimov, N. G. Leikov, A. V. Belov, I. M. Chertok, V. G. Kurt, A. M. Magun, and V. F. Melnikov, the Γ-I collaboration, AIP Conference Proceedings edited by J. Ryan and W. Vestrand, 29 (1993) 106-111.

[19] A.V. Belov and M.A. Livshits, Astr.Lett., 21 (1995) 37-40.

[20] D. F. Smart, M. A. Shea and K. O'Brien, Proceed 24th ICRC (Rome), 4 (1995) 171-174. H. Debrunner, J. A. Lockwood, C. Bradt, R. Buetikofer, J. P. Dezalay, E. Fleuckiger, A. Kuznetsov, J. M. Ryan, R. Sunyaev, O. V. Terekhov, G. Trottet, and N. Vilmar, Astrophys. J., 479 (1997) 997-1011.

[21] E. L. Chupp and J. M. Ryan, Research in Astron. Astrophys, 9 (2009) 11-40. See also the paper Rank et al, G. Rank, J. Ryan, H. Debrunner, M. McConnell, and V. Shoenfelder, A&A, 378 (2001) 1046-1066.

[22] K. R. Pyle and J. A. Simpson, Proceed. the 22nd ICRC (Dublin), 3 (1991), 53-56.

[23] Y. Muraki, K. Murakami, M. Miyazaki, K. Mitsui, S. Shibata, S. Sakakibara, T. Sakai, T. Takahashi, T. Yamada, and K. Yamaguchi, ApJL, 400 (1992) L75-L78.,Y. Muraki,

S. Sakakibara, S. Shibata, M. Satoh, K. Murakami, T. Takahashi, K. R. Pyle, T. Sakai, and K. Mitsui, J. Geomag. Geoelectr., 47 (1995) 1073-1078.

[24] A. Struminsky, M. Matsuoka, and K. Takahashi, ApJ, 429 (1994) 400-405.

[25] R. A. Schwartz, B. R. Dennis, G. J. Fishman, C.A. Meegan, R. B. Wilson, and W. S. Paciesas, the BATSE collaboration, Proceed. Of the Compton Observatory Science Workshop, NASA Conference Publication, 3137 (1992) 457-468.

[26] R. J. Murphy, G. H. Share, J. E. Grove, W. N. Johnson, R. L. Kinzer, R. A. Kroeger, J. D. Kurfess, M. S. Strickman, S. M. Matz, W. R. Purcell, M. P. Ulmer, D. A. Grabelsky, G. V. Jung, C. M. Jensen, D. J. Forrest and W. T. Vestrand, the OSSE collaboration, AIP Conference Proceedings edited by J. Ryan and W. Vestrand, 294 (1993) 15-20.

[27] E. J. Scheneid, K. T. S. Brazier, G. Kanbach, C von Montigny, H. A. Mayer-Hasselwander, D. L. Bertsch, C. E. Fichtel, R. C. Hartman, S. D. Hunter, D. J. Thompson, B. L. Dingus, P. Streekmar, Y. C. Lin, P. F. Micelson, P. L. Nolan, D. A. Kniffen, and J. R. Mattox, the EGRET collaboration, AIP Conference Proceedings edited by J. Ryan and W. Vestrand, 294 (1993) 94-99.

[28] S. Tsuneta and T. Naito, ApJ, 495 (1998) L67-L70.

[29] R. Ramaty and N. Mandzhaxvidze, AIP Conference Proceedings edited by J. Ryan and W. Vestrand, 294 (1993) 26-44.

[30] N. Mandzhavidze and R. Ramaty, ApJ, 389 (1992) 739-755.

[31] H. E. Petschek, Proceed. of the AAS-NASA symposium held on 28-30 October 1963 at Goddard Space Flight Center, (1964) p 425.

[32] S. Masuda, T. Kosugi, H. Hara, S. Tsuneta, and Y. Ogawara, Nature, 371 (1994) 495-497.

[33] K. Shibata, Astrophysics and Space Science, 264 (1999) 129-144., T. Yokoyama and K. Shibata, ApJL, 494 (1998) L113-116.

[34] G. D. Holman, AIP Conference Proceedings edited by R. Ramaty and N. Mandzhavidze, 206 (2000) 135-144.

[35] G. Rank, J. Ryan, H. Debrunner, M. McConnell, and V. Shönfelder, A&A, 378 (2001) 1046-1066.

[36] T. Sako, K. Watanabe, Y. Muraki, Y. Matsubara, H. Tsujihara, M. Yamashita, T. Sakai, S. Shibata, J. F. Valdés-Galicia, L. X. González, A. Hurtado, O. Musalem, P. Miranda, N. Martinic, R. Ticona, A. Velarde, F. Kakimoto, S. Ogio, Y. Tsunesada, H. Tokuno, Y. T. Tanaka, I. Yoshikawa, T. Terasawa, Y. Saito, T. Mukai, and M. Gros, ApJL 651 (2006) L69-L72.

[37] Y. Muraki, Proceed. 30[th] ICRC (Merida, Mexico) edited by R. Caballero, J. C. D'Olivo, G. Medina-Tanco and José F. Valdés-Galicia, 6 (2009) 181-194.

[38] Y. T. Tanaka, IAU telegram #3417 (for the flare of June 7, 2011) and #3886 (for the flare of January 23, 2012). Y. T. Tanaka, Paper presented at ICRC (Beijing) (2011) paper number #683 (for the flare of March 7, 2011).

[39] Lev Dorman, "Solar Neutrons and Related Phenomena, Astrophysics and Space Science Library Series 365", Springer (2010).

[40] H. Matsumoto, T. Goka, K. Koga, S. Iwai, T. Uehara, O. Sato, and S. Takagi, radiation Measurements, 33 (2001), 321-333., T. Doke, Science Asahi, 9 (1997) 38-43.

[41] H. Koshiishi, H. Matsumoto, A. Chishiki, T. Goka, and T. Omodaka, Radiation Measurement, 42 (2007) 1510-1520.

[42] Hinode web site: http://solar-b.nao.ac.jp/index_e.shtml

[43] Solar Dynamical Observatory web site; http://sdo.gsfc.nasa.gov/assets/img/browse/2011/06/07/20110607_062513_1024_0171.jpg For the Reference of RHESSI, see the paper by Vilmer,N., MacKinnon, A. L., & Hurford, G. J., "Properties of Energetic Ions in the Solar Atmosphere from gamma-Ray and Neutron Observations", Space Science Reviews, 159, (2011) 167-224.

[44] E. O. Flückiger, R. Bütikofer, A. Chlingalian, G. Hovspyan, Y. Muraki, Y. Matsubara, T. Sako, H. Tsuchiya and T. Sakai, Proceed. 27[th] ICRC, (2001) 3044-3047.

[45] Y. Muraki, H. Tsuchiya, K. Fujiki, S. Mausda, Y. Matsubara, Y. Menjo, T. Sako, K. Watanabe, M. Ohnishi, A. Shiomi, M. Takita, T. Yuda, Y. Katayose, N. Hotta, S. Ozawa, T. Sakurai, Y. H. Tan, and J. L Zhang, Astropart. Phys., 28 (2007) 119-131.

[46] J. F. Valdés-Galicia, Y. Muraki, H. Tsujihara, T. Sako, O. Musalem, A. Hurtado, L. X. González, Y. Matsubara, K. Watanabe, N. Hirano, N. Tateiwa,, S. Shibata, and T. Sakai, Nucl. Inst. Meth. A, 535 (2004) 656-664., L. X. González, F. Sánchez, J. F. Valdés-Galicia, Nucl. Inst. Meth. A, 613 (2010) 263-271.

[47] I. Imaida, Y. Muraki, K. Masuda, H. Tsuchiya, T. Hoshida, T. Sako, T. Koi, P. V. Ramanamurthy, T. Goka, H. Matsumoto, T. Omoto, A. Takase, K. Taguchi, I. Tanaka, M. Nakazawa, M. Fujii, T. Kohno, and H. Ikeda, Nucl. Inst. Meth. A, 421 (1999) 99-112

[48] K. Koga, T. Goka, H. Matsumoto, Y. Muraki, K. Masuda and Y. Matsubara, Radiation Measurements, 33 (2001), 287-291., K. Koga, T. Goka, H. Matsumoto, and Y. Muraki, Proceed. 31st ICRC (Lodz, Poland), 831 (2009) SH1.3.

[49] K. Koga, T. Goka, H. Matsumoto, T. Obara, Y. Muraki, and T. Yamamoto, Proceed. 21st ECRS (Kosice, Slovakia) edited by K. Kudela, P. Kiraly, and A. Wolfendale, (2008) 199-200.

[50] K. Koga, T. Goka, H. Matsumoto, T. Obara, Y. Muraki, and T. Yamamoto, Astrophys. Space Sci. Trans., 7 (2011) 411-416.

[51] K. Watanabe, private communication. See also the Phd thesis, Proceed. of the Cosmic-Ray Research of Nagoya University, Vol. 46 No. 2 (2005) 1-249, in English.

[52] Y. Muraki, Y. Matsubara, S. Masuda, S. Sakakibara, T. Sako, K. Watanabe, R. Bütikofer, E. O. Flückiger, A. Chlingalian, G. Hovspyan, F. Kakimoto, T. Terasawa, Y. Tsunesada, H. Tokuno, A. Velarde, P. Evenson, J. Poirier and T. Sakai, Astroparticle Physics, 29 (2008) 229-242.

[53] S. N. Kuznetsov, V. G. Kurt, I. N. Myagkova, B. Yu. Yushkov, and K. Kudela, Solar System Research, 40 No. 2 (2006) 104-110.

[54] D. Shiota, S. Tsuneta, M. Shimojo, N. Sakao, D. Orozco Suárez and R. Ishikawa, arXiV:1205.2154v1 [astro-ph.SR].

[55] Y. Hanaoka, Solar Physics, 173 (1997) 319-346.

**Table I** The list of solar flares with X-ray intensity exceeding X>2.0

| Date | Max. X-ray time | Class | Satellite position | Neutron existence |
|---|---|---|---|---|
| Feb. 6, 2010 | 18:59 | M2.9 | Sun side O | Neutron X |
| Feb. 7, 2010 | 02:34 | M6.4 | Sun side O | Neutron X |
| Feb. 8, 2010 | 07:43 | M4.0 | Sun side O | Neutron X |
| Feb. 8, 2010 | 13:47 | M2.0 | Sun side O | Neutron X |
| Feb. 12, 2010 | 13:47 | M2.0 | Sun side O | Neutron X |
| Feb. 13, 2011 | 17:38 | M6.6 | Eclipse X | Neutron X |
| Feb. 15, 2011 | 01:44 | X2.2 | Eclipse X | Neutron X |
| Feb. 18, 2011 | 10:11 | M6.6 | Sun side O | Neutron ? |
| Feb. 24, 2011 | 07:35 | M3.5 | Eclipse X | Neutron X |
| Mar. 7, 2011 | 20:12 | M3.7 | Sun side O | Neutron O |
| Mar. 8, 2011 | 10:44 | M5.3 | Eclipse X | Neutron X |
| Mar. 8, 2011 | 18:28 | M4.4 | Eclipse X | Neutron X |
| Mar. 9, 2011 | 23:23 | X1.5 | Eclipse X | Neutron X |
| Jun. 7, 2011 | 06:30 | M2.5 | O→X | Neutron O |
| Jul. 30, 2011 | 02:09 | M9.3 | Sun side O | Neutron ? |
| Aug. 3, 2011 | 13:48 | M6.0 | Sun side O | Neutron X |
| Aug. 4, 2011 | 03:57 | M9.3 | Sun side O | Neutron X? |
| Aug. 8, 2011 | 18:10 | M3.5 | Sun side O | Neutron ? |

| Aug. 9, 2011 | 03:54 | M2.5 | Sun side O | Neutron ? |
|---|---|---|---|---|
| Aug. 9, 2011 | 08:05 | X6.9 | Sun side O | Neutron X |
| Sep. 6, 2011 | 01:50 | M5.3 | Sun side O | Neutron X |
| Sep. 6, 2011 | 22:20 | X2.1 | Sun side O | Neutron X |
| Sep. 7, 2011 | 22:38 | X1.8 | Sun side O | Neutron X? |
| Sep. 8, 2011 | 15:46 | M6.7 | Sun side O | Neutron O |
| Sep. 9, 2011 | 06:11 | M2.7 | O→X | Neutron X |
| Sep. 22, 2011 | 11:01 | X1.4 | O→X | Neutron ? |
| Sep. 24, 2011 | 09:40 | X1.9 | X→O | Neutron X |
| Sep. 24, 2011 | 13:20 | M7.1 | Sun side O | Neutron X |
| <span style="color:red">Sep. 24, 2011</span> | <span style="color:red">19:18</span> | <span style="color:red">M3.0</span> | <span style="color:red">Sun side O</span> | <span style="color:red">Neutron O</span> |
| Sep. 24, 2011 | 20:36 | M5.8 | Sun side O | Neutron X |
| Sep. 25, 2011 | 02:33 | M4.4 | Eclipse(X→O) | Neutron ? |
| Sep. 25, 2011 | 04:50 | M7.4 | Eclipse(X→O) | Neutron ? |
| Sep. 25, 2011 | 15:33 | M3.7 | Sun side O | Neutron X |
| Sep. 26, 2011 | 05:08 | M4.0 | Sun side O | Neutron X |
| Sep. 26, 2011 | 14:45 | M2.6 | Sun side O | Neutron X |
| Oct. 2, 2011 | 00:50 | M3.9 | Sun side O | Neutron X |
| Nov. 2, 2011 | 22:01 | M4.3 | Sun side O | Neutron O |
| <span style="color:red">Nov. 3, 2011</span> | <span style="color:red">20:27</span> | <span style="color:red">X1.9</span> | <span style="color:red">Sun side O</span> | <span style="color:red">Neutron O</span> |
| Nov. 5, 2011 | 03:35 | M3.7 | Sun side O | Neutron X |
| Dec. 25, 2011 | 18:16 | M4.0 | Eclipse X | Neutron X |
| Jan. 19, 2012 | 16:05 | M3.2 | Sun side X→O | Neutron X |
| <span style="color:red">Jan. 23, 2012</span> | <span style="color:red">03:59</span> | <span style="color:red">M8.7</span> | <span style="color:red">Sun side O→X</span> | <span style="color:red">Neutron O</span> |
| Jan. 27, 2012 | 18:37 | X1.7 | Sun side O | Neutron X |
| Mar. 2, 2012 | 17:46 | M3.3 | Sun side O | Neutron ? |
| Mar. 4, 2012 | 10:45 | M2.0 | Eclipse X | Neutron X |
| Mar. 5, 2012 | 04:05 | X1.1 | X→O | Neutron ? |
| Mar. 7, 2012 | 00:24 | X5.4 | O→X | Neutron X |
| Mar. 9, 2012 | 03:45 | M6.3 | Sun side O | Neutron ? |
| Mar. 10, 2012 | 17:50 | M8.4 | Eclipse X | Neutron X |
| Mar. 13, 2012 | 17:25 | M7.8 | Sun side O | Neutron ? |
| May. 9, 2012 | 12:32 | M4.7 | Sun side O | Neutron X |
| May. 9, 2012 | 21:05 | M4.1 | Sun side O | Neutron X |
| May. 10, 2012 | 04:18 | M5.7 | Sun side O | Neutron X |
| May. 17, 2012 | 01:47 | M5.1 | Sun side O | Neutron X |
| Jul. 2, 2012 | 10:50 | M5.6 | Eclipse X | Neutron X |
| Jul. 2, 2012 | 20:05 | M3.8 | Eclipse X | Neutron X |
| Jul. 4, 2012 | 09:55 | M5.3 | Sun side O | Neutron X |
| Jul. 4, 2012 | 22:05 | M4.6 | Sun side O | Neutron X |
| Jul. 5, 2012 | 11:40 | M6.1 | Sun side O | Neutron X |
| Jul. 6, 2012 | 23:05 | X1.1 | X→O | Neutron X |
| Jul. 8, 2012 | 16:30 | M6.9 | Eclipse X | Neutron X |

| Jul. 17, 2012 | | 17:15 | M1.7 | Sun side O | Neutron X |
|---|---|---|---|---|---|
| Jul. 19, 2012 | | 05:58 | M7.7 | Sun side O | Neutron X |

**Table I:** The first to third columns of Table I correspond to the event date, maximum time of X-ray intensity, and flare size respectively. The fourth column indicates the ISS location regarding whether on the shadow or sunny side of Earth (X or ○). The fifth column indicates whether solar neutrons are involved in the data. The ○→ X and X→○ notations in the fourth column indicate that the ISS was moving from the sunny side to the shadow side or vice versa 30 minutes from the peak flare time. The ? mark of the fifth column indicates a possible neutron event with statistical significance of less than $3\sigma$ .

**Table II**

| ISS SEDA-FIB neutron event | | | Hinode | RHESSI | Fermi-LAT |
|---|---|---|---|---|---|
| Mar. 7, 2011 | M3.7 | 54, 35, 28 | × | ○ | Δ |
| Jun. 7, 2011 | M2.5 | 86, 2, 20 | ○ | ○ | ○ |
| Sep. 24, 2011 | M3.0 | 27, 2, 20 | ○ | ○ | Δ |
| Nov. 3, 2011 | X1.9 | 42, 15, 8 | Δ | × | × |
| Jan. 23, 2012 | M8.7 | 50, 14, 20 | ○ | × | × |

**Table II.** A score table between each satellite. The numbers of solar neutron events, from first to third, represent protons induced by neutrons, the incidents of which are consistent with those coming from the Sun, low-energy events difficult to identify as solar neutrons and ambiguous events due to low energy respectively. The marks ○, Δ, and × imply that each satellite was passing over the day side (○), partial day side (Δ) or night side (×) of Earth respectively. The event observed on September 8, 2011 was not involved in this Table, because it was observed near the SAA.

**Table III**

| Date of events | | | Total number of events | Average trigger rate (Hz) | Mean flux [events/cm$^2$sec] |
|---|---|---|---|---|---|
| Mar. 7, 2011 | M3.7 | SEDA | 89 | 0.09 | 0.042 |
| Jun. 7, 2011 | M2.5 | SEDA | 88 | 0.06 | 0.029 |
| Jun. 21, 1980 | X2.5 | SMM | 2,000 | 2.0 | 0.13 |

**Table III.** A comparison of solar neutron events observed by the SEDA-FIB detector with SMM event. The average trigger rate is simply obtained by dividing the total number of events by the observation time, while the mean flux is obtained taking the detection efficiency of both detectors into account (0.021 and 0.3 for SEDA and SMM) and also the detector area (100 and 450 cm$^2$ for SEDA and SMM).